\documentclass[conference]{IEEEtran}
\usepackage[UglyObsolete]{diagrams}
\usepackage{amsmath}
\usepackage{amssymb}
\usepackage{graphics}
\usepackage{latexsym}
\newtheorem{Main}{Main Lemma}
\newtheorem{proposition}{Proposition}

\newtheorem{definition}{Definition}
\newtheorem{example}{Example}
\newtheorem{algorithm}{Algorithm}

\author{
\IEEEauthorblockN{Hajime Matsui}
\IEEEauthorblockA{Toyota Technological Institute\\
Hisakata, Tenpaku, Nagoya, Japan\\
Email: matsui@toyota-ti.ac.jp}
}

\begin{document}
\title{Decoding a Class of Affine Variety Codes\\with Fast DFT}
\maketitle

\begin{abstract}
An efficient procedure for error-value calculations based on fast discrete Fourier transforms (DFT) in conjunction with Berlekamp-Massey-Sakata algorithm for a class of affine variety codes is proposed.
Our procedure is achieved by multidimensional DFT and linear recurrence relations from Grobner basis and is applied to erasure-and-error decoding and systematic encoding.
The computational complexity of error-value calculations in our algorithm improves that in solving systems of linear equations from error correcting pairs for many cases.
A motivating example of our algorithm in case of Reed-Solomon codes and a numerical example of our algorithm in case of a Hermitian code are also described.
\end{abstract}

\IEEEpeerreviewmaketitle

\section{Introduction}
Affine variety codes \cite{Fitzgerald-Lax98},\cite{Miura98} are the generalization of algebraic geometry (AG) codes and they are defined by a pair of a monomial order and any subset of $\mathbb{F}_q^N$, where $\mathbb{F}_q$ is the finite field of $q$-elements and $N$ is a positive integer.
It is known \cite{Fitzgerald-Lax98} that affine variety codes represent all linear codes.

The fast decoding of affine variety codes can be divided into two steps, namely error-location step and error-evaluation step.
As for the fast error-location step, it is shown \cite{generic06},\cite{BMS05} that Berlekamp--Massey--Sakata (BMS) algorithm finding the Gr\"obner basis of error-locator ideals for AG codes is generalized to that for affine variety codes.
As for the fast error-evaluation step, it is shown \cite{Little07} that Forney's method is generalized to affine variety codes.
On the other hand, in \cite{SITA11}, the author showed another fast error-evaluation method based on multidimensional DFT and linear recurrence relations from Gr\"obner basis of error-locator ideals.
The method in \cite{SITA11} is the generalization of the results in \cite{Sakata-Jensen-Hoholdt95},\cite{Sakata-erasure98} for AG codes and can be applied to erasure-and-error decoding and systematic encoding of affine variety codes.

In this paper, we restrict the codes to a subclass of affine variety codes that are defined by a pair of a monomial order and any subset of $\left(\mathbb{F}_q^\times\right)^N$, where $\mathbb{F}_q^\times=\mathbb{F}_q\backslash\{0\}$, in order to apply efficiently fast DFT.
This enables us to reduce the computational complexity of the error-evaluation step of \cite{SITA11} in their erasure-and-error decoding and systematic encoding.

The rest of this paper is organized as follows.
Sec.\ \ref{Motivation} explains the case of Reed--Solomon codes.
Sec.\ \ref{Main lemma} describes a lemma.
Subsec.\ \ref{Fourier-type} formulates multidimensional DFT.
Subsec.\ \ref{vector spaces} defines two vector spaces via Gr\"obner basis.
Subsec.\ \ref{Extension map} defines linear recurrence relations from Gr\"obner basis.
Subsec.\ \ref{Isomorphic map} gives an isomorphism between the vector spaces.
Sec.\ \ref{Application} applies the lemma to a class of affine variety codes.
Subsec.\ \ref{arbitrary subset} constructs the codes.
Subsec.\ \ref{Erasure-and-error} proposes an erasure-and-error decoding algorithm.
Subsec.\ \ref{Systematic encoding} relates erasure-only decoding with systematic encoding.
Sec.\ \ref{Estimation} estimates the computational complexity of our algorithm.

\section{A motivating example: Reed--Solomon codes
\label{Motivation}}

Throughout this paper, $\mathbb{N}_0$ is the set of non-negative integers and $\alpha$ is a fixed primitive element of finite field $\mathbb{F}_q$, where $q$ is a prime power.
Recall the encoding of Reed--Solomon (RS) codes by polynomial division
\begin{equation}\label{Euclidean}
c(x)=h(x)-R(x)=Q(x)G(x),\quad\deg(R)<n-k,
\end{equation}
where $h(x)=\sum_{0\le i<n}h_ix^i$ is an information polynomial with $h_i=0$ for $0\le i<n-k$; $G(x)=(x-1)\cdots(x-\alpha^{n-k-1})$, the generator polynomial; $R(x)$, the remainder of the division with quotient $Q(x)$.
Then $\left(c_i\right)_{0\le i<n}$ from $c(x)=\sum_{0\le i<n}c_i x^i$ is a codeword of the RS code
$$
C(m)=\left\{
\left(c_i\right)_{0\le i<n}\in \mathbb{F}_{q}^{n}\,\left|\,c(\alpha^i)=0\mbox{ for all }0\le i\le m
\right.\right\}
$$
with $n=q-1$ and $m=n-k-1$.
This method is {\it systematic}, i.e., $c_i=h_i$ for $n-k\le i<n$.
It is well-known that a {\it non-systematic} encoding method is given by the IDFT
$c_i=h(\alpha^{-i})=\sum_{0\le l<n}h_l\alpha^{-il}$ since $\left(c_i\right)_{0\le i<n}$ is another codeword of $C(m)$ by the Fourier inversion formula
\begin{equation}\label{inversion}
c(\alpha^{i'})=\sum_{0\le l<n}h_l\sum_{0\le i<n}\alpha^{i(i'-l)}=(q-1)h_{i'}.
\end{equation}

It is possible to encode systematically by adding another procedure ``extension.''
From \eqref{Euclidean}, we obtain $h(\alpha^i)=R(\alpha^i)$ for $0\le i<n-k$.
Define a vector $\left(d_i\right)_{0\le i<n}$ inductively by
\begin{equation}\label{extension encoding}
d_i=\left\{\begin{array}{ll}
h(\alpha^i)&0\le i<n-k\\
-\sum_{l=0}^{n-k-1}G_ld_{l+i-(n-k)}&
n-k\le i<n,
        \end{array}\right.
\end{equation}
where $G(x)=\sum_{l=0}^{n-k-1}G_lx^l+x^{n-k}$.
Then $d_i=R(\alpha^i)$ holds not only for $0\le i<n-k$ but also for $n-k\le i<n$; when $r$ denotes $n-k$, this fact is proved inductively by
\begin{equation}\label{proof}
d_{i}=-\sum_{l'=0}^{r-1}R_{l'}\alpha^{l'(i-r)}
\sum_{l=0}^{r-1}G_l\alpha^{l'l}
=\sum_{l'=0}^{r-1}R_{l'}\alpha^{l'i},
\end{equation}
where $R(x)$ is denoted by $R(x)=\sum_{l=0}^{r-1}R_lx^l$ and the last equality is deduced from $G(\alpha^{l'})=\sum_{l=0}^{r-1}G_l\alpha^{l'l}+\alpha^{l'r}=0$ for $0\le l'<r$.
Thus, the IDFT $\left(-d(\alpha^{-i})\right)_{0\le i<n}$ for $d(x)=\sum_{l=0}^{n-1}d_lx^l$ agrees with $(R_{i})_{0\le i<n}$ of $R(x)$ by using Fourier inversion formula \eqref{inversion};
$c(x)=h(x)-R(x)$ again indicates the encoding,
and moreover we obtain two ways of calculating $R(x)$, i.e., a 
remarkable commutative diagram.
\def\ext{\mathrm{Extension}}
\def\rest{\mathrm{Remainder}}
\def\comp{\mathrm{DFT}}
\def\Fti{\mathrm{IDFT}}
\begin{diagram}
\left(d_i\right)_{0\le i<n} & & \rTo^\Fti & & R(x) \\
\uTo^\ext & &  & & \uTo_\rest \\
\left(h(\alpha^i)\right)_{0\le i<r} & & \lTo^\comp & & h(x) \\
\end{diagram}

This encoding method is applicable to decoding method.
If we have received a polynomial $\overline{c}(x)=\sum_{0\le i<n}\overline{c}_ix^i=c(x)+e(x)$ containing an error polynomial $e(x)$ in the channel, the syndrome vector $\left(e(\alpha^i)\right)_{0\le i<n-k}$ can be computed as $\left(\overline{c}(\alpha^i)\right)_{0\le i<n-k}$ by substituting the roots of $G(x)$ into $\overline{c}(x)$.
Similarly to \eqref{extension encoding}, we define a vector $\left(s_i\right)_{0\le i<n}$ inductively by
\begin{equation}\label{extension decoding}
s_i=\left\{\begin{array}{ll}
\overline{c}(\alpha^i) & 0\le i<n-k\\
-\sum_{l=0}^{t-1}\sigma_ls_{l+i-(n-k)}&
n-k\le i<n,
        \end{array}\right.
\end{equation}
where $\sigma(x)=\sum_{l=0}^{t-1}\sigma_lx^l+x^t$ is the error-locator polynomial and $t\le(n-k)/2$ is assumed.
Then, it follows from the same argument as \eqref{proof} that $s_i=e(\alpha^i)$ with $e(x)=\sum_{l=0}^{n-1}e_lx^l$ for $0\le i<n$.
Thus, the IDFT $\left(-s(\alpha^{-i})\right)_{0\le i<n}$ for $s(x)=\sum_{l=0}^{n-1}s_lx^l$ agrees with $(e_{i})_{0\le i<n}$ by using Fourier inversion formula \eqref{inversion};
$c(x)=\overline{c}(x)-e(x)$ indicates the correct codeword.

\begin{example}\rm
Consider $C(3)$ of length $n=10$ and dimension $k=6$ over $\mathbb{F}_{11}=\mathrm{GF}(11)=\{0,1,2,\cdots,10\;\mathrm{mod}\,11\}$.
We fix a primitive element $2$ of $\mathbb{F}_{11}$.
We set information
$$\left(h_i\right)_{0\le i<10}=(0,0,0,0,1,7,3,2,0,5).$$
Then we have a non-systematic codeword
$$\left(c_i=h(2^{-i})\right)_{0\le i<10}=(7,10,3,5,6,1,5,1,3,3)\in C(3).$$
For systematic encoding, divide $h(x)$ by $G(x)=(x-1)(x-2)(x-4)(x-8)=9+x+4x^2+7x^3+x^4$ and obtain $h(x)=Q(x)G(x)+(9+2x+7x^3)$.
Thus $c(x)=h(x)-(9+2x+7x^3)$ is a systematic codeword, where
$$
\left(c_i\right)_{0\le i<10}=(2,9,0,4,1,7,3,2,0,5).
$$
This codeword is also computed by our method as follows.
\begin{align*}
\left(h(2^i)\right)_{0\le i<10}&=(7,3,3,1,5,1,6,5,3,10)\\
\left(d_i\right)_{0\le i<10}&=(7,3,3,1,3,0,4,4,6,4)\\
\left(-d(2^{-i})\right)_{0\le i<10}&=(9,2,0,7,0,0,0,0,0,0)\\
\left(c_i\right)_{0\le i<10}&=(2,9,0,4,1,7,3,2,0,5)\in C(3)
\end{align*}

Next, a received word can be decoded as follows.
\begin{align*}
\left(\overline{c}_i\right)_{0\le i<10}&=(1,9,0,4,1,7,3,2,0,10)\\
\left(\overline{c}(2^i)\right)_{0\le i<10}&=(4,7,3,1,2,6,4,7,5,4),\,(\sigma_l)=(6,4,1)\\
\left(s_i\right)_{0\le i<10}&=(4,7,3,1,0,5,2,6,8,9)\\
\left(-s(2^{-i})\right)_{0\le i<10}&=(10,0,0,0,0,0,0,0,0,5)\\
\left(c_i\right)_{0\le i<10}&=(2,9,0,4,1,7,3,2,0,5)\in C(3)\;\;\Box
\end{align*}
\end{example}

It is shown \cite{Magnetics09} that the encoding and decoding with extension and fast DFT and IDFT actually have less computational complexity than the ordinary ones for many cases of RS codes.

\section{Main lemma
\label{Main lemma}}

\subsection{Discrete Fourier transforms on $\left(\mathbb{F}_q^\times\right)^N$
\label{Fourier-type}}

Let $N$ be a positive integer and let
\begin{align*}
A&=\left\{\underline{a}=\left(a_1,\cdots,a_N\right)\left|
\begin{array}{c}
0\le a_i\le q-2\\
\mbox{for all }1\le i\le N
\end{array}\right\}\right.,\\
\Omega&=\left(\mathbb{F}_q^\times\right)^N=\left\{\left.\underline{\omega}=\left(\omega_1,\cdots,\omega_N\right)\,\right|\,\omega_1,\cdots,\omega_N\in\mathbb{F}_q^\times\right\},
\end{align*}
where $\mathbb{F}_q^\times=\left.\mathbb{F}_q\right\backslash\{0\}$.
Then $A$ is considered as the additive group $\left(\mathbb{Z}/(q-1)\mathbb{Z}\right)^N$, where $\mathbb{Z}/(q-1)\mathbb{Z}$ is the ring of integers modulo $(q-1)$.
In this subsection, the discrete Fourier transforms are defined as maps between two vector spaces, which are isomorphic to $\mathbb{F}_q^{|A|}=\mathbb{F}_q^{|\Omega|}$,
\begin{align*}
V_A&=\left\{\left(h_{\underline{a}}\right)_A\,\left|\,\underline{a}\in A,\,h_{\underline{a}}\in\mathbb{F}_q\right.\right\},\\
V_{\Omega}&=\left\{\left.\left(c_{\underline{\omega}}\right)_{\Omega}\,\right|\,\underline{\omega}\in\Omega,\,c_{\underline{\omega}}\in\mathbb{F}_q\right\}.
\end{align*}

\begin{definition}\label{substitution}
For $\left(c_{\underline{\omega}}\right)_\Omega\in V_\Omega$, let $\mathcal{F}\left(\left(c_{\underline{\omega}}\right)_\Omega\right)\in V_A$ be defined as
\begin{equation}\label{DFT}
\mathcal{F}\left(\left(c_{\underline{\omega}}\right)_\Omega\right)=\left(\sum_{\underline{\omega}\in\Omega}c_{\underline{\omega}}\underline{\omega}^{\underline{a}}\right)_A\in V_A,
\end{equation}
where $\underline{\omega}^{\underline{a}}=\omega_1^{a_1}\cdots\omega_N^{a_N}$.
The linear map $\mathcal{F}:V_\Omega\to V_A$ of \eqref{DFT} is called DFT on $\left(\mathbb{F}_q^\times\right)^N$.
Moreover, for $\left(h_{\underline{a}}\right)_A\in V_A$, let $\mathcal{F}^{-1}\left(\left(h_{\underline{a}}\right)_A\right)\in V_\Omega$ be defined as
\begin{equation}\label{IDFT}
\mathcal{F}^{-1}\left(\left(h_{\underline{a}}\right)_A\right)=\left((-1)^N\sum_{\underline{a}\in A}h_{\underline{a}}\underline{\omega}^{-\underline{a}}\right)_{\Omega}\in V_{\Omega}.
\end{equation}
The linear map $\mathcal{F}^{-1}:V_A\to V_\Omega$ of \eqref{IDFT} is called IDFT on $\left(\mathbb{F}_q^\times\right)^N$.
\hfill$\Box$
\end{definition}

Then we have Fourier inversion formulae:
Two linear maps $\mathcal{F}:V_\Omega\to V_A$ and $\mathcal{F}^{-1}:V_A\to V_\Omega$ are inverse each other, i.e., $\mathcal{F}^{-1}\left(\mathcal{F}\left(\left(c_{\underline{\omega}}\right)_\Omega\right)\right)=\left(c_{\underline{\omega}}\right)_\Omega$ and $\mathcal{F}\left(\mathcal{F}^{-1}\left(\left(h_{\underline{a}}\right)_A\right)\right)=\left(h_{\underline{a}}\right)_A$.

\subsection{Two vector spaces $V_S$ and $V_\Psi$
\label{vector spaces}}

Let $\Psi\subseteq\Omega$ and $n=|\Psi|$.
One of the two vector spaces in the lemma is given by
$$
V_\Psi=\left\{\left.\left(c_{\underline{\psi}}\right)_\Psi\,\right|\,\underline{\psi}\in\Psi,\,c_{\underline{\psi}}\in\mathbb{F}_q\right\}.
$$
The other of the two vector spaces is somewhat complicated to define, since it requires Gr\"obner basis theory \cite{Cox97}.
Let $\mathbb{F}_q[\underline{x}]$ be the ring of polynomials with coefficients in $\mathbb{F}_q$ whose variables are $x_1,\cdots,x_N$.
Let $Z_\Psi$ be an ideal of $\mathbb{F}_q[\underline{x}]$ defined by
$$
Z_\Psi=\left\{\left.f(\underline{x})\in\mathbb{F}_q[\underline{x}]\,\right|\,f(\underline{\psi})=0\mbox{ for all }\underline{\psi}\in\Psi\right\}.
$$
We fix a monomial order $\preceq$ of $\left\{\left.\underline{x}^{\underline{s}}\,\right|\,\underline{s}\in\mathbb{N}_0^N\right\}$ \cite{Cox97}.
We denote, for $f(\underline{x})\in\mathbb{F}_q[\underline{x}]$,
\begin{align*}
&\mathrm{LM}(f)=\max_{\preceq}\left\{\left.\underline{x}^{\underline{s}}\,\right|\,\underline{s}\in\mathbb{N}_0^N,\,f_{\underline{s}}\not=0\right\}
\\
&\mbox{if }\;f(\underline{x})=\sum_{\underline{s}\in\mathbb{N}_0^N,\,f_{\underline{s}}\not=0}f_{\underline{s}}\underline{x}^{\underline{s}}\in\mathbb{F}_q[\underline{x}]\;\mbox{ and }f(\underline{x})\not=0,
\end{align*}
where $\underline{x}^{\underline{s}}=x_1^{s_1}\cdots x_N^{s_N}$ for $\underline{s}=\left(s_1,\cdots,s_N\right)\in\mathbb{N}_0^N$.
Then the support $S=S_\Psi\subseteq\mathbb{N}_0^N$ of $Z_\Psi$ for $\Psi$ is defined by
\begin{equation}\label{delta set}
S=S_\Psi=\left.\mathbb{N}_0^N\right\backslash\left\{\mathrm{mdeg}\left(\mathrm{LM}(f)\right)\,\left|\,0\not=f(\underline{x})\in Z_\Psi\right.\right\},
\end{equation}
where $\mathrm{mdeg}\left(\underline{x}^{\underline{s}}\right)=\underline{s}\in\mathbb{N}_0^N$.
Then the other of the two vector spaces is given by
$$
V_S=V_{S_\Psi}=\left\{\left.\left(h_{\underline{s}}\right)_S=\left(h_{\underline{s}}\right)_{S_\Psi}\,\right|\,\underline{s}\in S_\Psi,\,h_{\underline{s}}\in\mathbb{F}_q\right\}.
$$
Since $\left.\left\{\underline{x}^{\underline{s}}\,\right|\,\underline{s}\in S_\Psi\right\}$ is a basis of quotient ring $\mathbb{F}_{q}[\underline{x}]/Z_\Psi$ viewed as a vector space over $\mathbb{F}_{q}$, $V_S$ is isomorphic to $\mathbb{F}_{q}[\underline{x}]/Z_\Psi$.
It is known \cite{{Fitzgerald-Lax98}} that the evaluation map
\begin{equation}\label{evaluation}
\mathbb{F}_{q}[\underline{x}]/Z_\Psi\ni
f\left(\underline{x}\right)\longmapsto
\left(f\left(\underline{\psi}\right)\right)_\Psi
\in V_\Psi
\end{equation}
is an isomorphism between two vector spaces.
Thus the map \eqref{evaluation} is also written as
\begin{equation}\label{ev}
V_S\ni
\left(h_{\underline{s}}\right)_S\longmapsto
\left(\sum_{\underline{s}\in S}h_{\underline{s}}\underline{\psi}^{\underline{s}}\right)_\Psi
\in V_\Psi,
\end{equation}
which is denoted as $\mathrm{ev}:V_S\to V_\Psi$.
In particular, it follows from the isomorphism \eqref{evaluation} or \eqref{ev} that $\left|S_\Psi\right|=|\Psi|$ and $\dim_{\mathbb{F}_q}V_S=\dim_{\mathbb{F}_q}V_{\Psi}=n$.

\subsection{Extension map $\mathcal{E}:V_S\to V_A$
\label{Extension map}}

Let $\mathcal{G}_\Psi$ be a Gr\"obner basis of $Z_\Psi$ with respect to $\preceq$.
Assume that $\mathcal{G}_\Psi$ consists of $d$ elements $\{g^{(w)}\}_{0\le w<d}$, where
\begin{equation}\label{grobner}
\begin{split}
&g^{(w)}=g^{(w)}(\underline{x})=\\
&\underline{x}^{\underline{s}_w}+\sum_{\underline{s}\in S_\Psi}g_{\underline{s}}^{(w)}\underline{x}^{\underline{s}}\in\mathbb{F}_q[\underline{x}]
\;\mbox{ with }\underline{s}_w\in\left.\mathbb{N}_0^N\right\backslash S_\Psi.
\end{split}
\end{equation}
For $\underline{a},\underline{b}\in A$, denote $\underline{a}\ge\underline{b}$ if $a_i\ge b_i$ for all $1\le i\le N$, or equivalently, if there is $\underline{c}\in A$ such that $\underline{a}=\underline{b}+\underline{c}$.

\begin{definition}\label{recurrence}
We define that $\left(k_{\underline{a}}\right)_{A}\in V_A$ is the extension of $\left(h_{\underline{s}}\right)_S\in V_S$ if and only if, for all $\underline{a}\in A$ and all $0\le w<d$,
\begin{equation}\label{generation}
k_{\underline{a}}=\left\{
\begin{array}{ll}
h_{\underline{s}} & \underline{a}=\underline{s}\in S_\Psi\\
-\sum_{\underline{s}\in S_\Psi}
g_{\underline{s}}^{(w)}k_{\underline{a}+\underline{s}-\underline{s}_w} & \underline{a}\ge\underline{s}_w.
\end{array}\right.\;\;\Box
\end{equation}
\end{definition}
Namely, each $k_{\underline{a}}$ for $\underline{a}\in A\left\backslash S_{\Psi}\right.$ satisfies at least one linear recurrence relation from \eqref{grobner} in $\mathcal{G}_\Psi$.
In fact, there is one-to-one correspondence between $\left(h_{\underline{s}}\right)_S\in V_S$ and $\left(k_{\underline{a}}\right)_{A}\in V_A$ that is the extension of some $\left(h_{\underline{s}}\right)_S\in V_S$; from a given $\left(h_{\underline{s}}\right)_S$, generate $(k_{\underline{a}})_A$ inductively by \eqref{generation}, where at least one $0\le w<d$ can be chosen such that $\underline{a}\ge\underline{s}_w$ and the resulting values do not depend on the choice and order of the generation because of the minimal property of Gr\"obner bases.

\begin{definition}\label{functions}
Denote $\mathcal{E}$ as the extension map via \eqref{generation}
$$
\mathcal{E}:V_S\rightarrow V_A\;\left[V_S\ni\left(h_{\underline{s}}\right)_S\mapsto\left(k_{\underline{a}}\right)_A\in V_A\right].$$
Moreover, denote $\mathcal{R}$ as the restriction map
$$
\mathcal{R}:V_\Omega\rightarrow V_\Psi\;\left[V_\Omega\ni\left(c_{\underline{\omega}}\right)_{\Omega}\mapsto\left(c_{\underline{\psi}}\right)_{\Psi}\in V_\Psi\right].
$$
Finally, denote $\mathcal{I}$ as the inclusion map
$$
\mathcal{I}:V_\Psi\rightarrow V_\Omega\;\left[V_\Psi\ni\left(c_{\underline{\psi}}\right)_{\Psi}\mapsto\left(c_{\underline{\omega}}\right)_{\Omega}\in V_\Omega\right],
$$
where $c_{\underline{\omega}}=c_{\underline{\psi}}$ if $\omega=\psi\in\Psi$ and $c_{\underline{\omega}}=0$ if $\omega\not\in\Psi$.
\hfill$\Box$
\end{definition}

\begin{proposition}\label{extension}
Let $\left(h_{\underline{s}}\right)_S\in V_S$.
Suppose that there is $\left(\epsilon_{\underline{\psi}}\right)_\Psi\in V_\Psi$ such that $\left(h_{\underline{s}}\right)_S=\left(\sum_{\underline{\psi}\in\Psi}\epsilon_{\underline{\psi}}\underline{\psi}^{\underline{s}}\right)_S$.
Moreover, let $\left(k_{\underline{a}}\right)_A=\mathcal{E}\left(\left(h_{\underline{s}}\right)_S\right)\in V_A$.
Then it follows that $\left(k_{\underline{a}}\right)_A=\left(\sum_{\underline{\psi}\in\Psi}\epsilon_{\underline{\psi}}\underline{\psi}^{\underline{a}}\right)_A$.
\hfill$\Box$
\end{proposition}
The proof of this proposition is similar to \eqref{proof}.

\subsection{Isomorphic map $\mathcal{C}:V_S\to V_\Psi$
\label{Isomorphic map}}

The following lemma is frequently used in this paper.
\begin{Main}
Let $\mathcal{G}_\Psi$ be a Gr\"obner basis of $Z_\Psi$ for $\Psi\subseteq\Omega$ and let $\mathcal{E}:V_S\to V_A$ be the extension map defined by \eqref{generation}.
Then we have
\begin{equation}\label{vanish}
\left(c_{\underline{\omega}}\right)_\Omega\in\mathcal{F}^{-1}\left(\mathcal{E}\left(V_S\right)\right)\:\Longrightarrow\:c_{\underline{\omega}}=0\,\mbox{ for all }\,\underline{\omega}\in\Omega\backslash\Psi.
\end{equation}
Moreover, the composition map $\mathcal{C}=\mathcal{R}\circ\mathcal{F}^{-1}\circ\mathcal{E}:V_S\to V_\Psi$ in the following commutative diagram
\def\ext{\mathcal{E}}
\def\rest{\mathcal{R}}
\def\comp{\mathcal{C}}
\def\Fti{\mathcal{F}^{-1}}
\begin{diagram}
V_A & & \rTo^\Fti & & V_{\Omega} \\
\uTo^\ext & &  & & \dTo_\rest \\
V_S & & \rTo^\comp & & V_\Psi \\
\end{diagram}
gives an isomorphism between $V_S$ and $V_\Psi$.
\hfill$\Box$
\end{Main}
Note that the first assertion \eqref{vanish} of the lemma deduces that $V_S$ is isomorphic to $V_{\Psi}$ by $\mathcal{C}$ since the image of $\mathcal{E}\left(V_S\right)$ by $\mathcal{F}^{-1}$ agrees with $V_\Psi$.

On the other hand, the inverse map $\mathcal{C}^{-1}:V_\Psi\to V_S$ of $\mathcal{C}$ can be written by
\begin{equation}\label{partial DFT}
V_\Psi\ni
\left(c_{\underline{\psi}}\right)_\Psi\longmapsto
\left(\sum_{\underline{\psi}\in\Psi}c_{\underline{\psi}}\underline{\psi}^{\underline{s}}\right)_S
\in V_S,
\end{equation}
which is the composition map $\mathcal{R}\circ\mathcal{F}\circ\mathcal{I}$, where $\mathcal{R}$ represents the restriction map $\mathcal{R}:V_A\rightarrow V_S\;\left[V_A\ni\left(c_{\underline{a}}\right)_A\mapsto\left(c_{\underline{s}}\right)_S\in V_S\right]$.
It is shown from the definitions that the matrices that represent two maps \eqref{ev} and \eqref{partial DFT} are transposed each other if the bases of vector spaces are fixed.

\section{Applications of main lemma
\label{Application}}

\subsection{Affine variety codes \cite{Fitzgerald-Lax98}
\label{arbitrary subset}}

Let $\Psi\subseteq\Omega$ and $R\subseteq S_\Psi$.
Consider two types of affine variety codes \cite{Fitzgerald-Lax98} with code length $n=|\Psi|$, where $\underline{\psi}^{\underline{r}}=\psi_1^{r_1}\cdots\psi_N^{r_N}$ is defined same as in \eqref{DFT}.
\begin{align}\label{code}
&C(R,\Psi)=\left\{\left(c_{\underline{\psi}}\right)_\Psi\in V_\Psi\left|
\begin{array}{c}
\sum_{\underline{r}\in R}h_{\underline{r}}\underline{\psi}^{\underline{r}}=c_{\underline{\psi}}\\
\mbox{for some }\left(h_{\underline{r}}\right)_R\in V_R
\end{array}
\right.\right\}\\
&C^\perp(R,\Psi)=\left\{\left(c_{\underline{\psi}}\right)_\Psi\in V_\Psi\left|
\begin{array}{c}
\sum_{\underline{\psi}\in\Psi}c_{\underline{\psi}}\underline{\psi}^{\underline{r}}=0\\
\mbox{for all }\underline{r}\in R
\end{array}
\right.\right\}\label{dual code}
\end{align}
It follows from the isomorphic map $\mathrm{ev}$ of \eqref{ev} that
\begin{equation}\label{ev image}
C(R,\Psi)=\mathrm{ev}\left(V_R\right)
\end{equation}
and that $\left.\left\{\left(\underline{\psi}^{\underline{r}}\right)_\Psi\,\right|\,\underline{r}\in R\right\}$ is a linearly independent basis of $C(R,\Psi)$.
Since $\sum_{\underline{\psi}\in\Psi}c_{\underline{\psi}}\underline{\psi}^{\underline{r}}$ in \eqref{dual code} is the value of the inner product for $\left(c_{\underline{\psi}}\right)_\Psi$ and $\left(\underline{\psi}^{\underline{r}}\right)_\Psi$ in $V_\Psi$, the dual code of $C(R,\Psi)$ is equal to $C^\perp(R,\Psi)$.
Thus the dimension or the number of information symbols $k$ of $C^\perp(R,\Psi)$ is equal to $n-|R|$, in other words, $n-k=|R|$.

Consider a subspace $V_{S\backslash R}$ of $V_S$ with $S=S_\Psi$.
Since $\mathcal{C}$ is isomorphic, we have $\mathcal{C}^{-1}\left(\left(c_{\underline{\psi}}\right)_\Psi\right)\in V_{S\backslash R}\Longleftrightarrow\left(c_{\underline{\psi}}\right)_\Psi\in \mathcal{C}\left(V_{S\backslash R}\right)$.
Thus we obtain
\begin{equation}\label{C image}
C^\perp(R,\Psi)=\mathcal{C}\left(V_{S\backslash R}\right),
\end{equation}
which is similar to \eqref{ev image}.
While the definition \eqref{dual code} of $C^\perp(R,\Psi)$ is indirect, the equality \eqref{C image} provides a direct construction and corresponds to non-systematic encoding of $C^\perp(R,\Psi)$.
Actually, non-systematic encoding is obtained as, for all $\left(h_{\underline{s}}\right)_S\in V_{S\backslash R}$, $\left(c_{\underline{\psi}}\right)_\Psi=\mathcal{C}\left(\left(h_{\underline{s}}\right)_S\right)\in C^\perp(R,\Psi)$ by \eqref{C image}.

\subsection{Erasure-and-error decoding
\label{Erasure-and-error}}

Suppose that erasure-and-error $\left(e_{\underline{\psi}}\right)_\Psi$ has occurred in a received word $\left(u_{\underline{\psi}}\right)_\Psi=\left(c_{\underline{\psi}}\right)_\Psi+\left(e_{\underline{\psi}}\right)_\Psi$ from a channel.
Let $\Phi_1\subseteq\Psi$ be the set of erasure locations and let $\Phi_2\subseteq\Psi$ be the set of error locations; we suppose that $\Phi_1$ is known but $\Phi_2$ and $\left(e_{\underline{\psi}}\right)_\Psi$ are unknown and that $e_{\underline{\psi}}\not=0\Leftrightarrow\underline{\psi}\in\Phi_1\cup\Phi_2$.
If $|\Phi_1|+2|\Phi_2|<d_\mathrm{FR}$ is valid, where $d_\mathrm{FR}$ is the Feng--Rao minimum distance bound \cite{Andersen-Geil08},\cite{BMS05}, then it is known that the erasure-and-error version \cite{Koetter98},\cite{Sakata-erasure98} of Berlekamp--Massey--Sakata (BMS) algorithm \cite{generic06},\cite{BMS05},\cite{Sakata-Jensen-Hoholdt95} calculates the Gr\"obner basis $\mathcal{G}_{\Phi_1\cup\Phi_2}$.
By using the recurrence from $\mathcal{G}_{\Phi_1\cup\Phi_2}$ and the lemma, the erasure-and-error decoding is realized as follows.

\begin{algorithm}{\it Finding erasure-and-errors}
\label{Finding erasure-and-errors}
\begin{description}
 \setlength{\itemsep}{1mm}
\item[Input:]\ $\Phi_1$ and a received word $\left(u_{\underline{\psi}}\right)_\Psi\in V_\Psi$
\item[Output:]\ \ $\left(c_{\underline{\psi}}\right)_\Psi\in C^{\perp}(R,\Psi)$
\item[Step 1.]\ $\left(v_{\underline{a}}\right)_A=\left(\sum_{\underline{\psi}\in\Phi_1}\underline{\psi}^{\underline{a}}\right)_A$
\item[Step 2.]\ Calculate $\mathcal{G}_{\Phi_1}$ from syndrome $\left(v_{\underline{r}}\right)_R$
\item[Step 3.]\ $\left(\widetilde{u}_{\underline{a}}\right)_{A}=\left(\sum_{\underline{\psi}\in\Psi}u_{\underline{\psi}}\underline{\psi}^{\underline{a}}\right)_A$
\item[Step 4.]\ Calculate $\mathcal{G}_{\Phi_1\cup\Phi_2}$ from $\left(\widetilde{u}_{\underline{r}}\right)_R$ and $\mathcal{G}_{\Phi_1}$
\item[Step 5.]\ $\left(e_{\underline{\psi}}\right)_\Psi=\mathcal{C}\left(\left(\widetilde{u}_{\underline{r}}\right)_R\right)$
\item[Step 6.]\ $\left(c_{\underline{\psi}}\right)_\Psi=\left(u_{\underline{\psi}}\right)_\Psi-\left(e_{\underline{\psi}}\right)_\Psi$
\hfill$\Box$
\end{description}
\end{algorithm}

\begin{example}\label{Hermitian decoding}
Consider a Hermitian code over $\mathbb{F}_9$ so that one can compare our methods with the conventional methods of algebraic geometry codes.
Putting $N=2$ and $q=9$, consider a monomial order $\preceq$ such that $(a,b)\preceq(a',b')\Leftrightarrow 3a+4b<3a'+4b'\mbox{ or }3a+4b=3a'+4b',\,a\le a'$ on $A=[0,7]^2$.
Choose $\Psi=\left\{\left.(\psi,\omega)\in\mathbb{F}_9^\times\,\right|\psi^4=\omega^3+\omega\right\}$.
In this case, the Gr\"obner basis $\mathcal{G}_{\Psi}$ consists of one $g(x,y)=x^4-y^3-y$ and the support $S_\Psi$ of $\mathcal{G}_{\Psi}$ is $\left\{\left.\left(s_1,s_2\right)\in A\right|s_2\le2\right\}$.
Let $R\subseteq S_\Psi$ be $R=\left\{\left.\left(r_1,r_2\right)\in S_\Psi\right|3r_1+4r_2\le11\right\}$ then $d_\mathrm{FR}=7$.
All values in error-only (i.e. $\Phi_1=\emptyset$) case of Algorithm \ref{Finding erasure-and-errors} are shown in bottom row of Fig.\ \ref{nonzero}, where the elements of $\mathbb{F}_9^\times$ are represented by their numbers of powers of a primitive element $\alpha$ with $\alpha^2+\alpha=1$, i.e., $0,1,\cdots,7$ means $\alpha^0,\alpha^1,\cdots,\alpha^7$, respectively, and $-1$ means $0\in\mathbb{F}_9$.
At Step 4 in Algorithm \ref{Finding erasure-and-errors}, the Gr\"obner basis $\mathcal{G}_{\Phi_2}$ of $Z_{\Phi_2}$ is obtained as
\begin{align*}
\mathcal{G}_{\Phi_2}=
\left\{
\begin{array}{l}
g^{(0)}=\alpha^2+\alpha^5 x+\alpha^6 y+x^2,\\
g^{(1)}=\alpha^5+x+\alpha^5 y+xy,\\
g^{(2)}=\alpha^4+y^2
\end{array}\right\}.
\quad\Box
\end{align*}
\end{example}

\subsection{Systematic encoding as erasure-only decoding
\label{Systematic encoding}}

Let $\Phi\subseteq\Psi\subseteq\Omega$ so that $\Phi$ corresponds to the set of redundant positions and $\Psi\backslash\Phi$ corresponds to the set of information positions.
Then $S_\Phi\subseteq S_\Psi$ holds since $Z_\Phi\supseteq Z_\Psi$ and the definition \eqref{delta set}.
From now on, consider the dual affine variety codes \eqref{dual code} with $R=S_\Phi$, i.e., $C=C^\perp(S_\Phi,\Psi)$.
Then $k=\dim_{\mathbb{F}_q}C=n-|\Phi|$ holds.
Systematic encoding means that, for a given information $\left(h_{\underline{\psi}}\right)_{\Psi\backslash\Phi}$, one finds $\left(c_{\underline{\psi}}\right)_\Psi\in C$ with $c_{\underline{\psi}}=h_{\underline{\psi}}$ for all $\underline{\psi}\in\Psi\backslash\Phi$.
Since $\Phi$ is known, systematic encoding can be viewed as erasure-only decoding for $\left(e_{\underline{\psi}}\right)_\Phi=\left(-c_{\underline{\psi}}\right)_\Phi$, i.e., $|\Phi_1|=|\Phi|$ and $|\Phi_2|=0$.
Actually, in case of RS codes, we viewed systematic encoding as erasure-only decoding since $|\Phi|=n-k=d_\mathrm{FR}-1$.
However, in general case, the correctable erasure-and-error bound $|\Phi_1|+2|\Phi_2|<d_\mathrm{FR}$ is not valid (cf. Example \ref{systematic encoding}).

Nevertheless we can show that systematic encoding works as erasure-only decoding.
We calculate the Gr\"obner basis $\mathcal{G}_\Phi$ in advance, which has the role of generator polynomials in case of RS codes.
Although the following Algorithm \ref{DFT systematic encoding} is equal to a special case of Algorithm \ref{Finding erasure-and-errors} for $\Phi_1=\Phi$ and $\Phi_2=\emptyset$, we assign it an algorithm in order to indicate the encoding.

\begin{algorithm}{\it DFT systematic encoding}
\label{DFT systematic encoding}
\begin{description}
 \setlength{\itemsep}{1mm}
\item[Input:]\ $\Phi$ and an information word $\left(h_{\underline{\psi}}\right)_{\Psi\backslash\Phi}$
\item[Output:]\ \ $\left(c_{\underline{\psi}}\right)_\Psi\in C$ with $\left(c_{\underline{\psi}}\right)_{\Psi\backslash\Phi}=\left(h_{\underline{\psi}}\right)_{\Psi\backslash\Phi}$
\item[Step 1.]\ $\left(\widetilde{u}_{\underline{a}}\right)_{A}=\left(\sum_{\underline{\psi}\in\Psi\backslash\Phi}h_{\underline{\psi}}\underline{\psi}^{\underline{a}}\right)_{A}$
\item[Step 2.]\ $\left(c_{\underline{\psi}}\right)_\Phi=-\mathcal{C}\left(\left(\widetilde{u}_{\underline{s}}\right)_{S_\Phi}\right)$
\hfill$\Box$
\end{description}
\end{algorithm}

\begin{figure*}[!t]
\centering
  \resizebox{15.5cm}{!}{\includegraphics{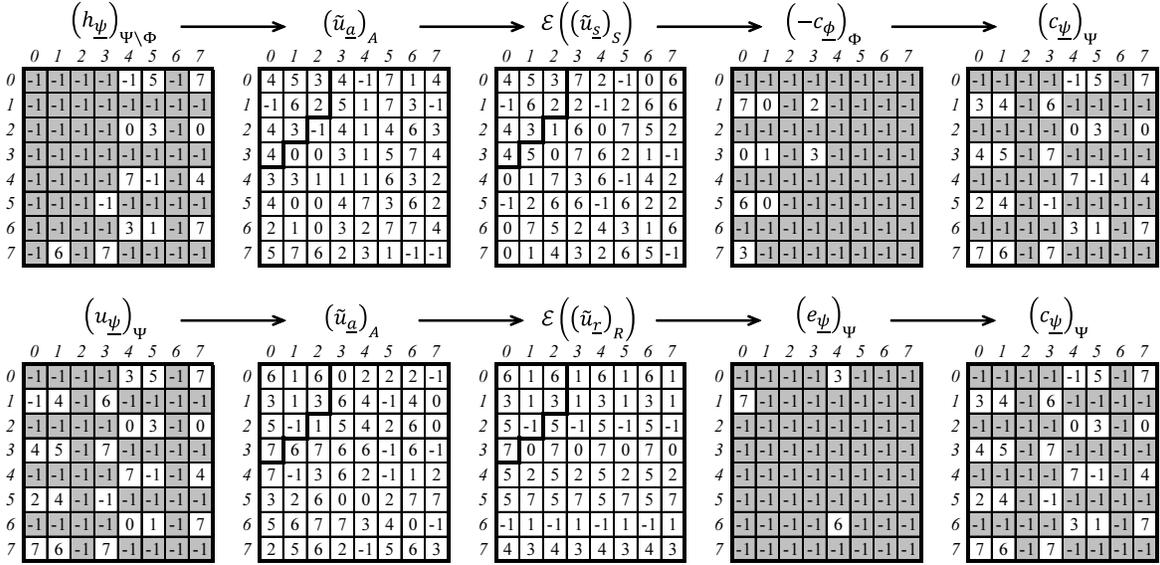}}
\caption{\small (Top row.) Systematic encoding of the shortened Hermitian code $C^\perp(S_{\Phi},\Psi)$ by Algorithm \ref{DFT systematic encoding}. A given information is $\left(h_{\underline{\psi}}\right)_{\Psi\backslash\Phi}$ and the systematic codeword is $\left(c_{\underline{\psi}}\right)_\Psi$. (Bottom row.) Its decoding by Algorithm \ref{Finding erasure-and-errors}. A received word is $\left(u_{\underline{\psi}}\right)_\Psi=\left(c_{\underline{\psi}}\right)_\Psi+\left(e_{\underline{\psi}}\right)_\Psi$.
\label{nonzero}}
\end{figure*}

\begin{example}\label{systematic encoding}
(Continued on Example \ref{Hermitian decoding}.)
Let
$$\Phi=\left\{
\begin{array}{cc}
(\alpha,1),
(\alpha,\alpha),
(\alpha,\alpha^3),
(\alpha^3,1),
(\alpha^3,\alpha),\\
(\alpha^3,\alpha^3),
(\alpha^5,1),
(\alpha^5,\alpha),
(\alpha^7,1)
\end{array}
\right\}.$$
Then Gr\"obner basis $\mathcal{G}_\Phi=\left\{g^{(0)},g^{(1)},g^{(2)},g^{(3)}\right\}$ is as follows:
\begin{align*}
&g^{(0)}=1+x^4,\quad g^{(3)}=1+y+y^3,\\
&g^{(1)}=\alpha\!+\!\alpha^2x\!+\!\alpha^3x^2\!+\!\alpha^4x^3\!+\!y(\alpha^5\!+\!\alpha^6x\!+\!\alpha^7x^2\!+\!x^3),\\
&g^{(2)}=\alpha^5\!+\!\alpha x\!+\!\alpha x^2\!+\!y(\alpha^7\!+\!\alpha^3x\!+\!\alpha^3x^2)\!+\!y^2(\alpha^4\!+\!x\!+\!x^2).
\end{align*}
Thus we have $S_\Phi=R$ and $|\Phi_1|=9>d_\mathrm{FR}=7$.
All values of Algorithm \ref{DFT systematic encoding} are shown in top row of Fig.\ \ref{nonzero}.
\hfill$\Box$
\end{example}

\section{Estimation of complexity
\label{Estimation}}

We estimate the number of finite-field operations, i.e., additions, subtractions, multiplications, and divisions, in Algorithm \ref{Finding erasure-and-errors} for codes \eqref{dual code}.
For the computation of DFT and IDFT, we employ multidimensional FFT algorithm; it is well-known that its number of finite-field operations is the order of $L\log L$, where $L$ is the data size and $L=q^N$ in our case.
Summarizing the results, we evaluate the algorithm as follows, where $n$ is code length, $N$ is dimension of $\Omega$, $q$ is finite-field size, and $d$ is the number of elements in Gr\"obner bases.
$$
{\renewcommand\arraystretch{1.0}
\begin{tabular}{|c|c|c|}
\hline
Algorithm \ref{Finding erasure-and-errors} & manipulation & order of bound \\
\hline
Step 1 & $\left(\sum_{\underline{\psi}\in\Phi_1}\underline{\psi}^{\underline{a}}\right)_A$ & $q^{N}\log q^{N}$\\
Step 2 & BMS & $dn^2$\\
Step 3 & $\left(\sum_{\underline{\psi}\in\Psi}u_{\underline{\psi}}\underline{\psi}^{\underline{a}}\right)_{A}$ & $q^{N}\log q^{N}$\\
Step 4 & BMS & $dn^2$\\
Step 5 & $\mathcal{C}\left(\left(\widetilde{u}_{\underline{r}}\right)_R\right)$ & $nq^{N}+q^{N}\log q^{N}$\\
Step 6 & $\left(u_{\underline{\psi}}\right)_\Psi-\left(e_{\underline{\psi}}\right)_\Psi$ & $n$\\
\hline
\end{tabular}}
$$

Thus the total number of finite-field operations in Algorithm \ref{Finding erasure-and-errors} is bounded by the order of $dn^2+nq^{N}$.
In particular, the number of finite-field operations for calculating error values apart from BMS algorithm has the order of $nq^{N}$.
In the proof \cite{Fitzgerald-Lax98} of $\{\mbox{linear codes}\}=\{\mbox{affine variety codes}\}$, $q^N$ is chosen as $q^{N-1}<n\le q^N$, which leads $nq^{N}<qn^2$.
Thus, if $q<n$, the computational complexity $qn^2$ of error-value calculation by Algorithm \ref{Finding erasure-and-errors} improves $n^3$ by error correcting pairs in \cite{Pellikaan92} and $n^3$ by inverse matrices of proper transforms in \cite{Saints95}.

\section*{Acknowledgment}
The author is grateful to Ryutaroh Matsumoto for his helpful comments and suggestions.
The author is also grateful to anonymous referees for their comments that were helpful.
This work was supported in part by KAKENHI, Grant-in-Aid for Scientific Research (C) (23560478).

\end{document}